# Quantum sum-frequency interference using up-converted photon pairs

**ATSUSHI SYOUJI[1,4,*], RYOSUKE SHIMIZU[1,2,5], SHIGEHIRO NAGANO[1], KOJI SUIZU[3], AND KEIICHI. EDAMATSU[1]**

[1]*Research Institute of Electrical Communication, Tohoku University, Sendai 980-8577, Japan*
[2]*PRESTO, Japan Science and Technology Agency (JST), Kawaguchi 332-0012, Japan*
[3]*Graduate School of Engineering, Nagoya University, Nagoya 464-8603, Japan*
[4]*Current address: Center for crystal science and technology, University of Yamanashi, Kofu 400-8511, Japan*
[5]*Current address: Graduate School of Informatics and Engineering, The University of Electro-Communications, 1-5-1 Chofugaoka, Chofu 182-8585, Japan*
*\*ashohji@yamanashi.ac.jp*

**Abstract:** We demonstrate a hybrid approach to the generation of photon pairs of a short wavelength with high brightness, by combining parametric down-conversion (SPDC) and up-conversion techniques. Photon pairs were generated at the wavelength of 1550 nm via SPDC, and converted to 516.7 nm through up-conversion with the pump at 775 nm. The quantum sum-frequency interference of the up-converted photon pairs exhibited a fringe period of 258.3 nm, which was 6 times shorter than the original wavelength, demonstrating that the energy-time correlation of the photon pairs was preserved. The technique simultaneously provides short fringe period beyond the classical limit and high brightness of the photon pairs.

## References and links

1. J. D. Franson, "Bell inequality for position and time," Phys. Rev. Lett. **62,** 2205 (1989).
2. J. G. Rarity, P. R. Tapster, E. Jakeman, T. Larchuk, R. A. Campos, M. C. Teich, and B. E. A. Saleh, "Two-photon interference in a Mach-Zehnder interferometer," Phys. Rev. Lett. **65,** 1348 (1990).
3. K. Edamatsu, R. Shimizu, and T. Itoh, "Measurement of the photonic de Broglie wavelength of entangled photon pairs generated by spontaneous parametric down-conversion," Phys. Rev. Lett. **89,** 213601 (2002).
4. A. N. Boto, P. Kok, D. S. Abrams, S. L. Braunstein, C. P. Williams, and J. P. Dowling, "Quantum interferometric optical lithography: Exploiting entanglement to beat the diffraction limit," Phys. Rev. Lett. **85,** 2733 (2000).
5. V. Giovannetti, S. Lloyd, and L. Maccone, "Quantum enhanced positioning and clock synchronization," Nature (London) **412,** 417 (2001).
6. Z.-Y. Zhou, S.-L. Liu, S.-K. Liu, Y.-H. Li, D.-S. Ding, G.-C. Guo, and B.-S. Shi, "Phase Measurement with Short-Wavelength NOON States by Quantum Frequency Up-Conversion," Phys. Rev. Appl. **7,** 064025 (2017).
7. M. W. Mitchell, J. S. Lundeen, and A. M. Steinberg, "Super-resolving phase measurements with a multiphoton entangled state," Nature **429,** 161(2004).
8. P. Walther, J. Pan, M. Aspelmeyer, R. Ursin, S. Gasparoni, and A. Zeilinger, "De Broglie wavelength of a non-local four-photon state," Nature **429,** 158 (2004).
9. T. Nagata, R. Okamoto, J. L. O'Brien, K. Sasaki, and S. Takeuchi, "Beating the standard quantum limit with four-entangled photons," Science **316,** 726 (2007).
10. I. Afek, O. Ambar, and Y. Silberberg, "High-NOON states by mixing quantum and classical light," Science **328,** 879 (2010).
11. K. J. Resch, K. L. Pregnell, R. Prevedel, A. Gilchrist, G. J. Pryde, J. L. O'Brien, and A. G. White, "Time-reversal and super-resolving phase measurements," Phys. Rev. Lett. **98,** 223601 (2007).
12. M. A. Albota and F. N. C. Wong, "Efficient single-photon counting at 1.55 µm by means of frequency upconversion," Opt. Lett. **29,** 1449 (2004).
13. H. Kamada, M. Asobe, T. Honjo, H. Takesue, Y. Tokura, Y. Nishida, O. Tadanaga, and H. Miyazawa, "Efficient and low-noise single-photon detection in 1550 nm communication band by frequency upconversion in periodically poled LiNbO$_3$ waveguides," Opt. Lett. **33,** 639 (2008).
14. C. Langrock, E. Diamanti, R. V. Roussev Y. Yamamoto M. M. Fejer, and H. Takesue, "Highly efficient single-photon detection at communication wavelengths by use of upconversion in reverse-proton-exchanged periodically poled LiNbO$_3$ waveguides," Opt. lett. **30,** 1725 (2005).

15. A. P.Vandevender, and P. G. Kwiat, "High efficiency single photon detection via frequency up-conversion," J. Mod. Opt. **51,** 1433 (2004).
16. A. Syouji, S. Nagano, R. Shimizu, K. Suizu, and K. Edamatsu, "Efficient up-conversion detection of 1550 nm photons using bulk periodically-poled LiNbO$_3$," Jpn. J. Appl. Phys. **49,** 040213 (2010).
17. S. Tanzilli, W. Tittel, M. Halder, O. Alibart, P. Baldi, N. Gisin, and H. Zbinden, "A photonic quantum information interface," Nature **437,** 116 (2005).
18. H. Takesue, "Erasing distinguishability using quantum frequency up-conversion," Phys. Rev. Lett. **101,** 173901 (2008).
19. P. Kumar, "Quantum frequency conversion," Opt. lett. **15,** 1476 (1990).
20. J. Huang, and P. Kumar, "Observation of quantum frequency conversion," Phys. Rev. Lett. **68,** 2153 (1992).
21. S. Nagano, M. Konishi, T. Shiomi, and M. Minakata, "Study on formation of small polarization domain inversion for high-efficiency quasi-phase-matched second-harmonic generation device," Jpn. J. Appl. Phys. **42,** 4334 (2003).
22. S. Nagano, R. Shimizu, Y. Sugiura, K. Suizu, K. Edamatsu, and H. Ito, "800-nm band cross-polarized photon pair source using type-II parametric down-conversion in periodically poled lithium niobate," Jpn. J. Appl. Phys. **46,** L1064 (2007).
23. L. E. Myers, R. C. Eckardt, M. M. Fejer, R. L. Byer, W. R. Bosenberg, and J. W. Pierce, "Quasi-phase-matched optical parametric oscillators in bulk periodically poled LiNbO$_3$," J. Opt. Soc. Am. B **12,** 2102 (1995).
24. J. Brendel, E. Mohler, and W. Martienssen, "Time-resolved dual-beam two-photon interferences with high visibility," Phys. Rev. Lett. **66,** 1142 (1991).
25. M. H. Rubin and Y. H. Shih, "Models of a two-photon Einstein-Podolsky-Rosen interference experiment," Phys. Rev. A **45,** 8138 (1992).
26. Y. H. Shih, A. V. Sergienko, M. H. Rubin, T. E. Kiss, and C. O. Alley, "Two-photon interference in a standard Mach-Zehnder interferometer," Phys. Rev. A **49,** 4243 (1994).
27. A. V. Burlakov, M. V. Chekhova, D. N. Klyshko, S. P. Kulik, A. N. Penin, Y. H. Shih, and D. V. Strekalov, "Interference effects in spontaneous two-photon parametric scattering from two macroscopic regions," Phys. Rev. A **56,** 3214 (1997).

## 1. Introduction

It is well-known that energy-time entangled photon pairs via spontaneous parametric down-conversion (SPDC) exhibit quantum interference with sum-frequency oscillation of the photons [1–3]. The *quantum sum-frequency interferometry* opens up the way to precise metrology beyond the classical limit and stimulates various applications, e.g., quantum lithography [4] and high-accuracy positioning systems [5]. Recently, higher order quantum sum-frequency interferometry for multi-photon states such as high-NOON states [6] has attracted much attention in order to obtain shorter interference fringe periods. Thus far, three- [7], four- [8, 9], five- [10] and six- [11] photon interference experiments have been reported. However, these experiments have suffered from low generation efficiencies. For example, in the experiment reported in Ref. [8], the four-photon counting rate was only 45 counts in 600 s. Although practical applications aiming at accurate positioning or high-resolution imaging beyond the classical limit require not only shorter fringe periods but also higher brightness, it is difficult to realize multi-photon states of high brightness because the generation and detection probabilities of multi-photons exponentially decreases with increasing number of photons concerned.

Recently, efficient wavelength conversion techniques have been developed and applied to single and a few photons. Such techniques are quite valuable for various quantum optical applications. Especially, up-conversion technique has been proven to be an efficient method to detect telecom wavelength photons used quantum information and communication technologies [12–16]. For example, we have reported the up-conversion experiment of 1550 nm photons to 516.5 nm with the internal conversion efficiency of 96 % [16]. In addition, it has been demonstrated that quantum coherence of the original photons is preserved in the up-converted photons [17–20]. Hence, we expect that, if we up-convert photon pairs generated by SPDC to a shorter wavelength, the up-converted photon pairs will exhibit quantum sum-frequency interference with much shorter fringe intervals than the original wavelength while keeping the brightness, i.e., the number of photons pairs concerned.

In this paper, we report our experiment on quantum sum-frequency interference using up-converted photon pairs. In our experiment, the photon pairs generated originally at 1550 nm are up-converted to 516.7 nm. We demonstrate that the quantum sum-frequency interference of the up-converted photons exhibits the fringe period of 258.3 nm, which is 6 times shorter, as expected, than the wavelength of the original photons. With the help of efficient up-conversion technique we developed [16], we achieved the two-photon coincidence rate of ~2000 counts/s.

## 2. Experimental setup

Our system exploits bulk periodically-poled LiNbO$_3$ (PPLN) crystals and a pulsed laser system [21, 22], as illustrated in Fig. 1. A cw laser with a wavelength $\lambda_1$ of 1550 nm from an external cavity laser diode was chopped by a LiNbO$_3$ intensity modulator (Fujitsu ES/FTM H74M-5208-J156) and then amplified by a two-stage Er-doped fiber amplifier (EDFA) [16]. The pulse width was 7.5 ns and the repetition rate was 4 MHz. The laser pulse was then fed to the second harmonic generation (SHG) crystal (SHG-PPLN) to get the second harmonics (SH) at $\lambda_2$ =775 nm. The SH was the pump source of both the SPDC and the up-conversion processes. Using SPDC crystal (SPDC-PPLN), we obtained SPDC photons with a wavelength $\lambda_1$ of 1550 nm. Then the SPDC photons were up-converted with the SH to a wavelength $\lambda_3$ of 516.7 nm. The PPLN crystals were fabricated using the full-cover electrode domain inversion method [23] in a z-cut congruent LiNbO$_3$ crystal (length = 38 mm, thickness = 0.5 mm). In our experiment, we used three PPLN crystals: one has a poling period of $\Lambda$ = 6.6 μm designed for the up-conversion (UPC-PPLN), i.e., sum-frequency generation in type-0 (eee) quasi phase matching, and the other two crystals have a period of 18.4 μm for second harmonic generation (SHG-PPLN) and SPDC (SPDC-PPLN).

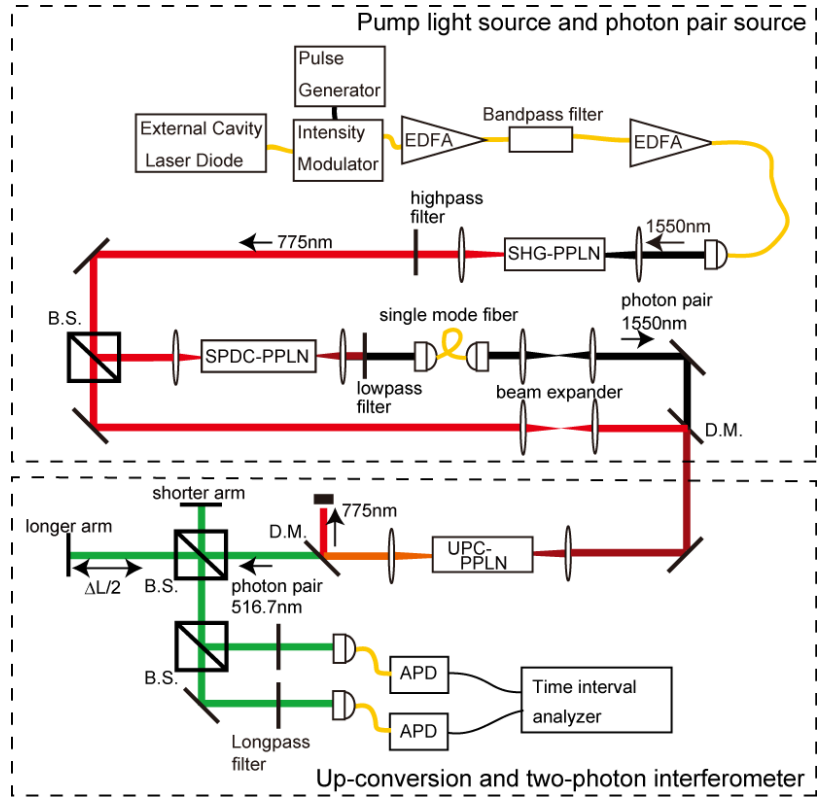

Fig. 1. Experimental setup. Upper: light source section. Lower: up-conversion and interferometer section to observe sum-frequency interference of up-converted photon pairs. D.M.: dichroic mirrors, B.S.: beam splitters, APD: avalanche photodiodes.

The SPDC photons (1550 nm wavelength) were generated by the PPLN crystal ($\Lambda$ = 18.4 µm). We obtained the average single photon counting rate 12600 counts/s and coincidence rate 1320 counts/s when the average pump power for the SPDC was 3 mW (0.75 nJ/pulse) and the pump pulse duration was 2.5 ns. From these values, we estimated that the average number of generated photon pairs was $4.0 \times 10^4$ counts/mW. After cleaning up the spatial modes of the SPDC photons by letting them pass through a 300-mm-long polarization maintaining single-mode fiber, we fed the photons into the PPLN crystal ($\Lambda$ = 6.6 µm) for the up-conversion process together with the pump pulse using a dichroic mirror. Here, we settled the average pump power to be 1800 mW (450 nJ/pulse) and the pulse duration to be 7.5 ns so as to obtain the maximum conversion efficiency of ~96 % [16]. The average pump power for the SPDC was settled to be 9 mW.

We then carried out quantum interference experiments using the up-converted photon pairs (516.6 nm wavelength). As shown in Fig. 1, the photon pairs were fed into an asymmetric Michelson interferometer, which served as a folded Franson interferometer [1, 24] along with the two-photon coincidence detection. The photons exited from the interferometer were detected by a pair of fiber-coupled avalanche photodiodes (APD, Micro Photon Devices PDMNPD5CTB, timing jitter = 50 ps), and the arrival time difference was analyzed by the time interval analyzer (ORTEC Model 9308-PCI, temporal resolution = 25 ps). In the Franson-type two-photon interference, it is crucial to distinguish the two propagation events inside the interferometer [24–26]: one is that each constituent photon separately propagates through the interferometer arms, and the other is that the two photons go together through either arm of the interferometer. In the former case, the difference $\Delta\tau_L$ of the arrival time between the two photons is given as $\Delta\tau_L = \pm\Delta L/c$, while the latter case causes no difference in the arrival time at the output port of the interferometer. Here $\Delta L$ is the optical path-length difference of the interferometer, c the speed of light. Thus, by setting a coincidence time window $|\Delta t| < \Delta\tau_L$, we selected the coincidence event in which both photons passed together through the same arms. A mirror in the longer arm was settled on a piezo stage so that the optical path-length difference $\Delta L$ was scanned by each 6 nm, and we plotted the detected number of the selected coincidence events as a function of $\Delta L$. In the ideal case, we can observe the two-photon's sum-frequency interference fringes with the interval of 516.6/2=258.3 nm and 100 % visibility [1, 25].

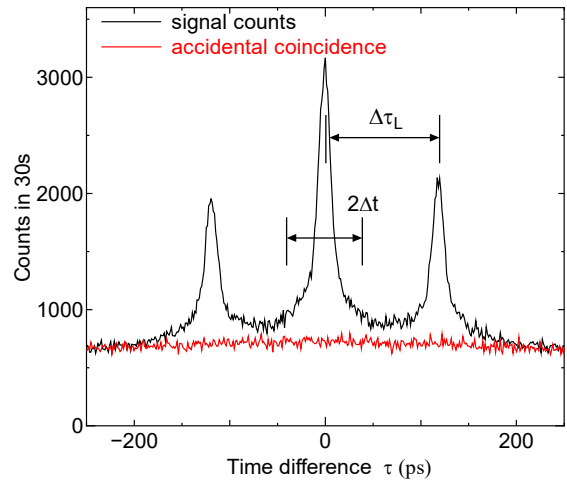

Fig. 2. Time correlation histogram of photon pairs. The peak distance of $\Delta\tau_L$ is obtained from arm length difference $\Delta L$ of asymmetric Michelson interferometer. $2\Delta t$ is time window for sum-frequency interference. The black line is the signal counts and the red line is accidental coincidence (see the text).

## 3. Results and Discussion

Figure 2 shows a resultant time correlation histogram of photons as a function of the detection time difference $\Delta\tau$ between two photons, when $\Delta L$ = 36 mm ($\Delta\tau_L = \pm120$ ps). In this histogram, we clearly see three peaks. The two side peaks at $\Delta\tau_L = \pm120$ ps correspond to the coincidence events in which one photon passed through the shorter arm and the other passed through the longer arm, while the central peak corresponds to the coincidence events in which both photons passed together through the same arm. Thus, we clearly see the same pair-wise property of the up-converted photons as in pairs generated by SPDC. Setting the coincidence time window to $|\Delta t| < 46$ ps, we can clearly separate the coincidence events corresponding to the central peak from those to the side peaks. With these events thus selected, we observe the two-photon's sum-frequency interference. The coherence time $\tau_{coh}$ for the two-photon's sum-frequency interference is essentially governed by the coherence time of the pump pulse for the SPDC [3, 25, 26]. Therefore, in our case, we expect that the two-photon coherence time $\tau_{coh}$ is almost identical to the pump pulse duration of 7.5 ns. Our experiment satisfies the condition of $|\Delta t| < \Delta\tau_L \ll \tau_{coh}$. The red line shown in fig. 2 is the accidental coincidence between the events separated by $\Delta\tau$ = 250 ns, i.e., the separation of adjacent pump pulses. The accidental coincidence includes three origins: accidentals between signal photons generated in different pairs, noise photons generated in the up-conversion process as discussed in [16], and the dark counts of the APDs. In the following, we subtract the contribution of accidental coincidence from the data.

Figure 3 (a) shows the observed interference pattern as a function of the path-length difference $\Delta L$. We clearly observe the fringe period exhibiting 258.3 nm with a visibility of 70 %. The fringe period is shortest ever been reported in two-photon interference experiments. Table 1 is a comparison of previous reports of multi-photon interference experiments with our study. We see that our technique has an advantage of reasonably high event rate (~2000 counts/s). Also, the event probability of our system is considerably higher than those in the past experiments.

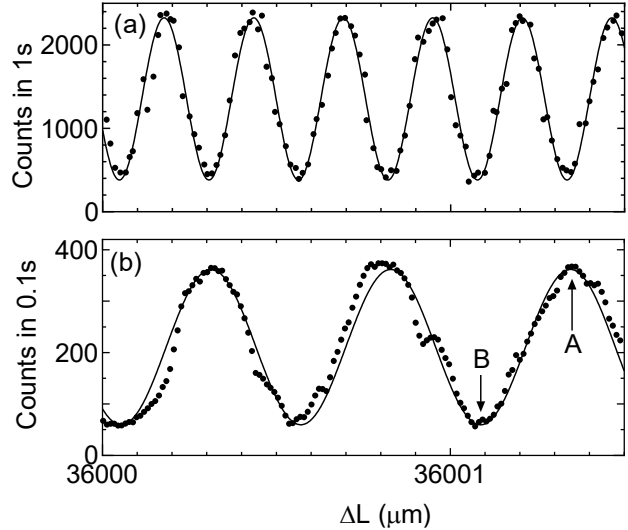

Fig. 3. (a): Two-photon interference of up-converted photon pair without accidental coincidence and dark counts. (b): classical interference of the up-converted laser pulse. "A" and "B" are a peak and a dip of the interference fringe, respectively.

**Table 1. Comparison of some reports of multi-photon interference experiments and our study. cw is continuous wave.**

|  | number of interfering photons | fringe period (nm) | event rate (counts/s) | pump repetition (MHz) | event probability (/pulse) |
|---|---|---|---|---|---|
| S. Tanzilli, *et al.* [17] | 2 | 448.6 | 8 | cw |  |
| P. Walther *et al.* [8] | 2 | 395 | 4800 | 76 | $6.4 \times 10^{-5}$ |
| T. Nagata *et al.* [9] | 4 | 195 | 0.04 | 77 | $5.2 \times 10^{-10}$ |
| P. Walther *et al* [8] | 4 | 197.5 | 0.06 | 76 | $7.7 \times 10^{-10}$ |
| I. Afek *et al.* [10] | 5 | 161.6 | 0.015 | 80 | $1.9 \times 10^{-10}$ |
| Our study | 2 | 258.3 | 2000 | 4 | $5 \times 10^{-4}$ |

Although we demonstrated the short fringe period in the two-photon interference using up-converted photon pairs, the interference visibility (~70 %) in our experiment was not satisfactory. In order to find out the origin of the visibility degradation, we checked conventional interference of classical light of 516.6 nm, which is up-converted from 1550 nm laser pulse. As shown in Fig. 3(b), the interference fringe with a period of 516.6 nm was observed as expected, but the visibility was only ~70 %, which was comparable to the visibility of the two-photon interference. In addition, the interference fringes contain noisy fluctuations suggesting that the phase of the signal light was not stable enough. To see this more in detail, we observed temporal profiles of the classical signal pulse using a pin-photodiode after the interferometer. The black line "A" and the red line "B" in Fig. 4 are the temporal profiles of pulses corresponding to the fringe positions indicated by "A" (at a bright fringe) and "B" (at a dark fringe) in Fig. 3(b), respectively. In "B", we see that the destructive interference at the dark fringe was not perfect, especially in the region of leading and trailing edges. By integrating and comparing the whole pulse area of "A" and "B", we estimate the interference visibility of 69 %, which is consistent with the observed fringe visibility (~70 %). If integrate the pulse area from 2.5 ns to 7.5 ns so as to eliminate the contribution from the edge regions, the estimated visibility becomes 82 %. From these results, we see that a part of degradation of the fringe visibility came from the incomplete interference at the pulse edge regions. As mentioned above, our laser pulse was generated by a LiNbO$_3$ intensity modulator. In general, such an intensity modulator consisting of an internal interferometer has an unavoidable fast phase swing during switching transition. This fast phase swing may cause degradation of coherence of transmitted light at the pulse edge regions. Hence, the incomplete interference appeared at the both edges probably arise from this property of signal light we used. As the same manner, the degradation of the two-photon interference visibility for the up-converted photon pairs would occur because the photon pairs inherit the phase information of the SPDC pump light [27]. The up-conversion process would also affect the coherence of light at the edge regions because we used the same pump light for up-conversion. Thus, when designing interferometric experiments using pulsed lasers generated by fast intensity modulation, one must take care of phase instability including fast phase swing at pulse edges. In addition, even though we could eliminate the edge regions, the estimated visibility (82 %) would be still imperfect. Although the reason for this is not clear at present, one possible reason is the imperfect spatial mode of the upconverted light.

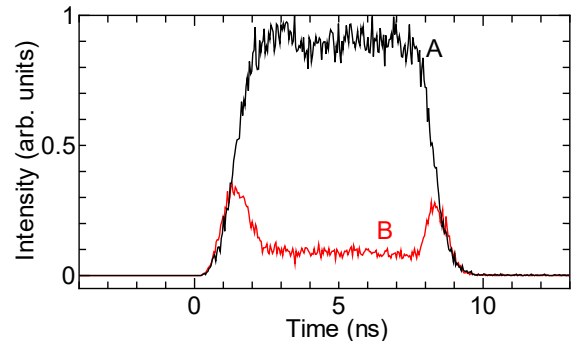

Fig. 4. Temporal profiles of up-converted laser pulse after interference. "A" and "B" are the signals detected at path-length difference positions of "A" and "B" in Fig. 3(b), respectively.

### 4. Conclusion

In summary, we have demonstrated the two-photon sum-frequency interference of up-converted photon pairs. Observed fringe period of 258.3 nm, which is 6 times shorter than the original wavelength of the photon, is the shortest ever been reported in two-photon interference experiments. The obtained event rate (~2000 counts/s) as well as the event probability is one of the highest reported in this kind of experiments, and can be further improved by increasing the pump repetition rate. Our technique will be useful in experiments that take advantage of quantum interference with shorter fringe periods with high event rates.

### Acknowledgments

This work was supported by a Grant-in-Aid for Creative Scientific Research (17GS1204) from the Japan Society for the Promotion of Science.